\documentclass[aps,pre,twocolumn,superscriptaddress,showpacs]{revtex4}
\usepackage{float}
\usepackage{graphicx}
\begin{document}

\title{Effective medium approach for stiff polymer networks with flexible cross-links}

\author{C.P. Broedersz}
\affiliation{Department of Physics and Astronomy, Vrije
Universiteit, 1081 HV Amsterdam, The Netherlands}
\author{C. Storm}
\affiliation{Department of Physics and Astronomy, Vrije
Universiteit, 1081 HV Amsterdam, The Netherlands}
\affiliation{Instituut Lorentz, Leiden University, P.O. Box 9506,
2300 RA Leiden, The Netherlands}\affiliation{Department of Applied
Physics and Institute for Complex Molecular Systems, Eindhoven
University of Technology, P.\ O.\ Box 513, NL-5600 MB Eindhoven, The
Netherlands}
\author{F.C. MacKintosh}
\email{fcm@nat.vu.nl} \affiliation{Department of Physics and
Astronomy, Vrije Universiteit, 1081 HV Amsterdam, The Netherlands}

\date{\today}

\begin{abstract}
Recent experiments have demonstrated that the nonlinear elasticity
of \emph{in vitro} networks of the biopolymer actin is dramatically
altered in the presence of a flexible cross-linker such as the
abundant cytoskeletal protein filamin. The basic principles of such
networks remain poorly understood. Here we describe an effective
medium theory of flexibly cross-linked stiff polymer networks. We
argue that the response of the cross-links can be fully attributed
to entropic stiffening, while softening due to domain unfolding can
be ignored. The network is modeled as a collection of randomly
oriented rods connected by flexible cross-links to an elastic
continuum. This effective medium is treated in a linear elastic
limit as well as in a more general framework, in which the medium
self-consistently represents the nonlinear network behavior. This
model predicts that the nonlinear elastic response sets in at
strains proportional to cross-linker length and inversely
proportional to filament length. Furthermore, we find that the
differential modulus scales linearly with the stress in the
stiffening regime. These results are in excellent agreement with
bulk rheology data.
\end{abstract}


\maketitle

\section{Introduction}\label{Introduction}
The mechanical response and locomotion of living
cells is mainly controlled by the cellular
cytoskeleton. The cytoskeleton is a highly
composite network of various stiff biopolymers,
along with various binding proteins for force
generation, cross-linking and polymer growth
regulation. Understanding the basic physics that
governs the mechanical properties of a composite
biopolymer network represents an important
biophysical challenge that will help elucidate
the mechanics of a living cell. In addition to
their importance for cell mechanics, cytoskeletal
networks have also demonstrated novel rheological
properties, especially in numerous \emph{in
vitro}
studies~\cite{JanmeyJCB91,MacReview97,Wirtz,Gittes97,Gardel04,Storm05,BauschNatPhys,Chaudhuri07,Janmey07,Kasza2007}.
However, there have been few theoretical or
experimental studies that address the composite
nature of the
cytoskeleton~\cite{Gardel06,Wagner06,DiDonna06,DischerLubensky,broedersz08,Karen}.
Recent experiments on F-actin networks with the
highly compliant cross-linker filamin, in
particular, have demonstrated several striking
features: These networks can have a linear
modulus as low as 1 Pa, which is significantly
lower than for actin gels with incompliant
cross-links, and yet they can withstand stresses
of 100 Pa or more and can stiffen dramatically by
up to a factor of 1000 under applied
shear~\cite{Gardel06,Kasza2007}. Both the linear
and nonlinear elastic properties of actin-filamin
gels appear to be dramatically affected by the
flexible nature of the cross-links, resulting in
novel behavior as compared to actin-networks with
incompliant cross-links, and to synthetic polymer
gels. This suggests new network design principles
that may be extended to novel synthetic materials
with engineered cross-links~\cite{Wagner06}.
However, the basic physics of networks with
flexible cross-links remain unclear.

\begin{figure}
\centering
\includegraphics[width=240 pt]{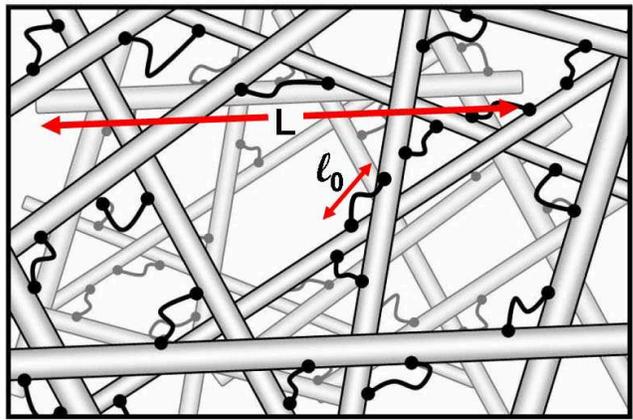}
\caption{Schematic figure of an isotropic stiff polymer network with
highly compliant cross-linkers.} \label{intro_networkschematic}
\end{figure}

In this article we provide a detailed description of an effective
medium approach to describe the nonlinear elastic properties of
composite networks consisting of stiff filaments linked by highly
flexible cross-links~\cite{broedersz08}. A schematic image of the
network we aim to model is shown in
Fig.~\ref{intro_networkschematic}. The network is composed of
randomly oriented filaments/rods of length $L$, which are linked
together by highly flexible cross-linkers. The cross-links consist
of two binding domains interconnected by a thermally fluctuating
flexible polymer chain of length $\ell_0$. The compliance of such a
cross-linker is entropic in nature.

Adopting the WLC model, we can fully characterize the cross-linkers
with a contour length $\ell_0$ and a persistence length
$\ell_p$\cite{Bustamante,Marko95}. The WLC force-extension curve,
which is shown in Fig.~\ref{EMA_HRschemeandx-linkforceextension}~c)
demonstrates the dramatic stiffening of the cross-linker as it
reaches its full extension. Indeed, atomic force microscope (AFM)
measurements show that an actin cross-linker such as filamin can be
accurately described as a wormlike chain
(WLC)~\cite{Schwaiger,Furuike01}. At large mechanical loads,
however, the experimental force-extension curve deviates
significantly from WLC behavior. The polymer chain in cross-linkers
such as filamin consists of repeated folded protein domains, which
unfold reversibly at sufficiently large mechanical loads. The
experiments by Furuike et al.~\cite{Furuike01} show that after an
initial stiffening regime at a force-threshold of $\approx
100~\text{pN}$ one of the protein domains unfolds reversibly. The
accompanied increase in contour length results in a strong decrease
in the cross-linkers stiffness. This softening is immediately
followed by WLC stiffening as the thermal undulations of the
lengthened cross-linker are stretched out. This leads to an elastic
response that alternates between entropic stiffening and softening
caused by domain unfolding, resulting in a sawtooth force-extension
curve.

It has been suggested that the unfolding behavior of filamin is
crucial for the mechanical properties of networks with such
cross-linkers\cite{Gardel06,Furuike01,DiDonna06}. Simulations of
stiff polymer networks, assuming a sawtooth force-extension curve
for the unfoldable cross-links, reveal that such networks exhibit a
fragile state in which a significant fraction of cross-linkers is at
the threshold of domain unfolding~\cite{DiDonna06}. This results in
strain softening of the network under shear, inconsistent with the
pronounced stiffening response observed experimentally in
actin-filamin gels~\cite{Gardel06,Kasza2007}. We estimate, however,
that under typical \emph{in vitro} experimental conditions, domain
unfolding in the cross-links is highly unlikely. For domain
unfolding to occur with multiple filamin crosslinks with tensions of
order $100~\text{pN}$, the resulting tension in the actin filaments
is likely to exceed rupture forces of order $300~\text{pN}$ of
F-actin~\cite{Tsuda96}. Also, a simple estimate of the macroscopic
stress corresponding to even a small fraction of filamins under
$100~\text{pN}$ tensions is larger than the typical limit of shear
stress before network failure is observed. Therefore, we do not expect
domain unfolding to occur. Rather, it seems likely that cross-link unbinding
occurs before sufficiently large sufficiently large forces are attained for a significant amount
of domain unfolding. Detailed estimates based on experiments
suggest filamin tensions only of order 1-5 pN at network
failure~\cite{Karen}. It has also been shown in single molecule
experiments~\cite{Ferrer08} that filamin unbinds from F-actin at
forces well below the forces required for unfolding, which indicates
that cross-linker unfolding is highly unlikely to occur in typical
network conditions. Therefore, we consider only the initial
stiffening of the cross-links, which we show can account well for
the observed nonlinear elasticity of actin-filamin gels.

Our model consists of a network of stiff filaments connected by
flexible cross-linkers. The compliance of such a network is expected
to be governed by the cross-linkers. The stiff filaments provide
connectivity to the network and constraint the deformation of the
cross-linkers, thereby setting the length scale of the effective
unit cell of the network. Consequently, we expect that the
elasticity of the network will be controlled by the filament length
$L$ and network connectivity, which is expressed in terms of the
number of cross-link per filament $n$. Therefore, we describe the
network with a model in which the basic elastic element consists of
a single stiff rod and many compliant cross-linkers that are
connected to a surrounding linear elastic medium.

\section{Effective Medium Approach}\label{Effective Medium Approach}
Networks of semiflexible polymers with point-like
incompliant cross-links have been studied
extensively~\cite{MacKPRL95,Gardel04,Storm05,Onck2005,Tharmann2007,Palmer2008}.
These systems exhibit two distinct elastic
regimes: One in which the deformation is affine
down to the smallest length scales of the network
and a regime that is characterized by highly
non-affine deformations.
Simulations~\cite{Head03,Wilhelm2003} have shown
that the deformation of these networks becomes
more affine with increasing cross-link
concentration and polymer length length, which
has been borne out by
experiments~\cite{Gardel04,Liu07}. The elastic
response of the network can fully be accounted
for by the stretching modes of the polymers in
the affine regime. In addition to stretching
modes, stiff polymers can also store energy in a
non-affine bending mode. Indeed, it has been
shown that in sparser networks, in which there
are fewer constraint on the constituting
polymers, non-affine bending modes dominate the
elastic
response~\cite{Head03,Wilhelm2003,Onck2005}. We
will, however, not consider the sparse network
limit here.

\begin{figure}
\centering
\includegraphics[width=240 pt]{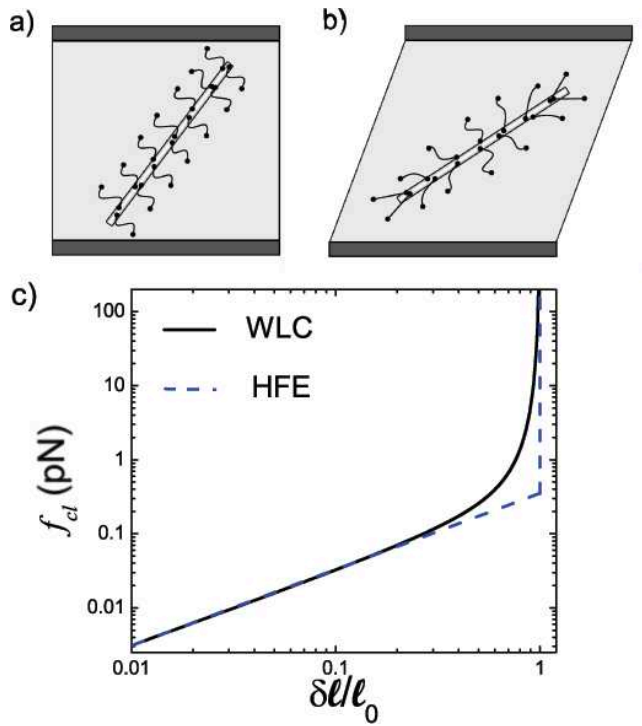}
\caption{(Color online) a) a single filament connected by $n$
flexible cross-links to the surrounding network, which we model as
an effective elastic continuum (shown here as a grey background) and
b) illustrates the proposed nonuniform deformation of the
cross-linkers on a single filament in a sheared background medium.
c) Force-extension curve of a Hookean Finite Extendable (HFE)
cross-linker (dashed blue curve) and of a WLC cross-linker (solid
black curve).} \label{EMA_HRschemeandx-linkforceextension}
\end{figure}

We expect the soft stretching modes of the cross-linkers to govern
the elasticity of a dense network of stiff polymers with highly
flexible cross-links. However, the large separation in size and
stiffness between cross-links and filaments does imply a non-uniform
deformation field for the cross-links at the sub-filament level.
On a coarse-grained level the network deforms affinely and stretches
the cross-links as depicted in
Fig.~\ref{EMA_HRschemeandx-linkforceextension} b). The network
surrounding this particular rod is shown here as a grey background.
The deformation of the cross-links increases linearly from $0$ in
the center towards a maximum value at the boundaries of the rod. At
small strains the cross-links are very soft and follow the
deformation of the stiffer surrounding medium. However, at a strain
$\gamma_c\sim\ell_0/L$ the outer-most cross-links reach their full
extension and, consequently, stiffen dramatically. This suggest the
existence of a characteristic strain $\gamma_c$, for the onset of
the nonlinear response of the network.

The macroscopic elasticity of the network results from the tensions
in all the constituting filaments. The tension in a particular
filament can be determined by summing up the forces exerted by the
cross-links on one side of the midpoint of the filament. We will
employ an effective medium approach to calculate these forces as a
function of filament orientation and the macroscopic strain. Thus,
we model the network surrounding one particular rod, as a continuum,
which effectively represents the elasticity of the network, as
depicted in Fig.~\ref{EMA_HRschemeandx-linkforceextension} a) and
b). We then proceed by considering contributions from rods over all
orientations to calculate the macroscopic response of the network.

The remainder of this article is organized as follows. First we
study a model in which the effective medium is treated as a linear
elastic continuum. In this model we will describe the cross-links
both as linear springs with finite extension, and also as WLC
cross-links. We analyze our model in both a fully 3D network, as
well as a simplified 1D representation, which already captures the
essential physics of the nonlinear behavior. At large strains, when
many of the cross-linkers are extended well into their nonlinear
regimes, it is no longer realistic to model the surrounding network
as a linear medium. Therefore, we extend our linear medium model in
a self-consistent manner, replacing the embedding medium by a
nonlinear effective medium whose elastic properties are determined
by those of the constituent rods and linkers. This self-consistent
model can quantitatively account for the nonlinear response found in
prior experiments on actin filamin networks~\cite{Gardel06,Karen}.
Finally, we show how we can compute the tension profiles along the
filaments and we demonstrate how to use these to express the
macroscopic stress in terms of the maximum force experienced by a
single cross-link.

\section{The Linear medium model}\label{Linear medium model}
We first develop a one dimensional representation of our model,
which will be used in section~\ref{3D network calculation} to
construct a more realistic three dimensional model. Also we will
restrict the treatment here to a linear description of the effective
medium, a constrain that we lift in section~\ref{Self-Consistent
Approach}.

Consider a rigid rod of length $L$ connected by $n$ flexible
cross-links to an elastic medium. We shall refer to such an elastic
unit as a Hairy Rod (HR). The medium is subject to an externally
imposed extensional strain $\epsilon$ parallel to the orientation of
the rod. The presence of the HR in the medium reduces the
deformation of the medium at a position $x$ in the rest frame of the
rod by an amount $u_{EM}(x, \epsilon) = \epsilon x -
u_{cl}(x,\epsilon)$, where $u_{cl}(x,\epsilon)$ is the extension of
a cross-linker at a distance $x$ from the center of the rod. The
magnitude of $u_{cl}(x,\epsilon)$ and $u_{EM}(x,\epsilon)$ are set
by requiring force balance between the cross-links and the medium.
\begin{equation}\label{LMM_forcebalance}
      f_{cl}(u_{cl}(x,\epsilon))=K_{EM} u_{EM}(x,\epsilon),
\end{equation}
where $f_{cl}(u)$ is the force-extension curve of a single
cross-linker. The tension $\tau_0$ in the center of the rod is found
by summing up the forces exerted by the stretched cross-links on one
side of the midpoint of the rod. Assuming a high, uniform line
density $n/L$ of cross-links along the rod, we can write the sum as
an integral
\begin{equation}\label{LMM_midpointtension}
      \tau_0(\epsilon)=\frac{n}{L} \int_0^{L/2} dx' \,
      f_{cl}\big(u_{cl}(x',\epsilon)\big).
\end{equation}
where $u_{cl}(x',\epsilon)$ is found by solving
Eqn.~(\ref{LMM_forcebalance}). The full tension profile
$\tau(\epsilon,x)$ is found by replacing the lower limit of the
integration by $x$
\begin{equation}\label{LMM_tensionprofile}
      \tau(\epsilon,x)=\frac{n}{L} \int_x^{L/2} dx' \,
      f_{cl}\big(u_{cl}(x',\epsilon)\big)
\end{equation}

\subsection{Hookean finite extendable cross-linkers}\label{simplecrosslinkers}
\begin{figure}
\centering
\includegraphics[width=240 pt]{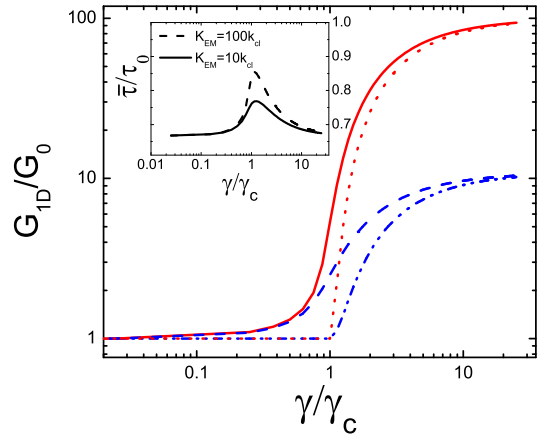}
\caption{(Color online) a) The modulus $G_{1D} = \tau_0/\epsilon$
for the 1D representation of the linear medium model with HFE
cross-links with $K_{EM} = 10 k_{cl}$ (blue dashed dotted curve) and
$K_{EM} = 100 k_{cl}$ (red dotted curve). We also show $G_{1D}$ for
the model with WLC cross-links with $K_{EM} = 10 k_{cl}$ (blue
dashed curve) and $K_{EM}= 100 k_{cl}$ (red solid curve). The inset
shows the ratio of the average tension $\bar\tau$ and the midpoint
tension $\tau_0$.} \label{LinMedModel}
\end{figure}

We can solve Eqns.~(\ref{LMM_forcebalance}) and
(\ref{LMM_midpointtension}) to compute the the midpoint tension in a
rod, as soon as a force-extension curve for the cross-links is
specified. In the absence of unfolding or unbinding, we can describe
the force-extension behavior of a flexible cross-linker such as
filamin with the WLC model, as depicted with the black solid line in
Fig.~\ref{EMA_HRschemeandx-linkforceextension}~c). It is instructive
to simplify the WLC force-extension curve by assuming a Hookean
response with a spring constant $k_{cl}$ up to an extension
$\ell_0$, which is the molecular weight of the cross-linker. The
spring constant $k_{cl}=\frac{2}{3}\frac{k_B T}{\ell_p\ell_0}$ is
found from the WLC model for small extensions in the limit $\ell_p
\ll \ell_0$~\cite{Marko95}, where $k_B T$ is the thermal energy.
Beyond an extension $\ell_0$, the cross-linker becomes infinitely
stiff. The force-extension curve of these Hookean Finite Extendable
(HFE) cross-links is shown as a blue dashed curve in
Fig.~\ref{EMA_HRschemeandx-linkforceextension}~c). The finite
extensibility of the cross-links implies a critical strain
$\epsilon_c=\frac{\ell_0}{L/2}$ at which the cross-linkers at the
boundaries of the rod reach full extension. For strains $\epsilon
\le \epsilon_c$
\begin{equation}\label{LMM_tensionsimple1}
    \tau_0(\epsilon)= \frac{n}{L} \int_0^{L/2} \, dx' \, \frac{k_{cl} K_{EM}}{k_{cl}+K_{EM}}\epsilon
    x'.
\end{equation}
Thus, the midpoint tension depends linearly on strain for $\epsilon
\le \epsilon_c$. For larger strains, the expression for the midpoint
tension in a hairy rod in Eq.~(\ref{LMM_midpointtension}) reads
\begin{eqnarray}\label{LMM_tensionsimple2}
    \tau_0(\epsilon)&=& \frac{n}{L} \int_0^{\ell_0/ \epsilon} \, dx' \, \frac{k_{cl} K_{EM}}{k_{cl}+K_{EM}}\epsilon x'\\
    &+&\frac{n}{L} \int_{\ell_0/ \epsilon}^\frac{L}{2} \, dx' \,
    \left[\frac{k_{cl} K_{EM}}{k_{cl}+K_{EM}}\ell_0+K_{EM}(\epsilon
    x'-\ell_0)\right]\nonumber.
\end{eqnarray}
The expression has separated into two integrals, clearly
representing a sum over the cross-links with an extension $<\ell_0$
and a sum over the cross-links that have already reached full
extension. We also note that beyond $\epsilon_c$ the midpoint
tension depends nonlinearly on strain. Using
Eq.~(\ref{LMM_tensionsimple2}) we compute the 1D modulus $G_{1D} =
\tau_0/\epsilon$, as shown in Fig.~\ref{LinMedModel}. Below the
critical strain, the response is dominated by the linear elasticity
of the cross-links $G_{1D} \approx \frac{1}{8} n k_{cl} L$. The
cross-links at the edge of the rod become rigid at a strain
threshold $\epsilon_c=2 \ell_0/L$. As the strain is further
increased, the outer cross-links stiffen consecutively, resulting in
a sharp increase of $G_{1D}$. At large strains, $G_{1D}$
asymptotically approaches a second linear regime $\sim\frac{1}{8} n
K_{EM} L$.

\subsection{Worm Like Chain Cross-Linkers}\label{LMM WLC Crosslinkers}
\begin{figure}
\centering
\includegraphics[width=240 pt]{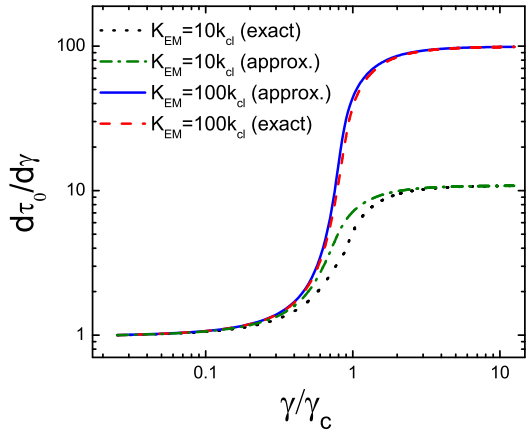}
\caption{(Color online) The 1D modulus $G_{1D}=\tau/\epsilon$ of the
rod with WLC cross-linkers as a function of the extensional strain
$\epsilon$ imposed on the medium parallel to the orientation of the
rod. The solid red curve shows the exact calculation using
Eqs.~(\ref{LMM_forcebalance}) and (\ref{LMM_midpointtension}) and
the dashed blue curve shows the approximate calculation using
Eq.~(\ref{LMM_tensionWLC}).} \label{LMM_exactapproxcomparison}
\end{figure}

We now consider flexible cross-linkers described by the more
realistic WLC force-extension curve, as depicted by the solid line
in Fig.~\ref{EMA_HRschemeandx-linkforceextension}~c). The
force-extension relation is well described by the interpolation
formula ~\cite{Marko95}
\begin{equation}\label{LMM_WLClinker definition}
   f_{cl}(u) = \frac{k_B T}{\ell_p} \left(\frac{1}{4\left(1-\frac{u}{\ell_0}
   \right)^2}-\frac{1}{4}+\frac{u}{\ell_0}\right),
\end{equation}
where $k_{B}T$ is the thermal energy. Using
Eqs.~(\ref{LMM_forcebalance}) and (\ref{LMM_midpointtension}) we can
calculate the 1D modulus $G_{1D}$ for cross-linkers with this
force-extension curve. The result of this calculation is shown in
Fig.~\ref{LMM_exactapproxcomparison}. The force-extension curve of
the WLC cross-linker is linear up to extensions very close to
$\ell_0$, upon which a pronounced stiffening occurs, as shown in
Fig.~\ref{EMA_HRschemeandx-linkforceextension}~c). We can exploit
this, together with the property that for a dense network the medium
is much stiffer than the flexible cross-linkers $K_{EM} \gg k_{cl}$
to write an approximate expression for the tension in a hairy rod in
a closed form analogous to Eq.~(\ref{LMM_tensionsimple2}).
\begin{eqnarray}\label{LMM_tensionWLC}
    \!\!\!&\tau_0(\epsilon)&= \frac{n}{L} \int_0^{\ell_0/ \epsilon} \!\!\!dx' \int_0^{\epsilon x'} du \frac{k_{cl}(u)
    K_{EM}}{k_{cl}(u)+K_{EM}}\\
  \!\!\!   \!\!\! &+& \!\!\!\frac{n}{L} \int_{\ell_0/ \epsilon}^\frac{L}{2} \!\!\! dx'
   \left[ \int_0^{\ell_0} \!\!\!du \frac{k_{cl}(u) K_{EM}}{k_{cl}(u)+K_{EM}}+K_{EM}(\epsilon
   x'-\ell_0)\right],\nonumber
\end{eqnarray}
where $k_{cl}(u)$ is the differential stiffness $d f_{cl}/du$ of the
WLC cross-linker. This equation states that an HR unit deforms
essentially affine up to the critical strain. Beyond $\epsilon_c$,
those cross-links that have reached full extension are no longer
compliant and start to pull back on the surrounding medium. The
approximate calculation of $d\tau_0/d\gamma$ using
Eq.~(\ref{LMM_tensionWLC}) is shown together with the exact
calculation performed with Eq.~(\ref{LMM_midpointtension}) in
Fig.~\ref{LMM_exactapproxcomparison}. This graph demonstrates that
the approximation captures the essential behavior of the exact
curve, and results only in a minor quantitative difference in the
cross-over regime. Therefore, we will continue constructing our
model using this approximation.

The 1D modulus calculated with Eq.~(\ref{LMM_tensionWLC}) is shown
for the WLC cross-links together with the results of the HFE
cross-links in Fig.~\ref{LinMedModel}. Although the main behavior is
very similar to that of the HFE cross-linker model, the use of the
more realistic WLC force-extension curve has introduced a
considerable smoothing of the cross-over. The nonlinear behavior in
the WLC force-extension curve initiates slowly well before full
extension, resulting in a more gradual onset of nonlinear behavior
of the HR with WLC cross-linkers. Remarkably, the characteristic
strain $\epsilon_c$ for the nonlinear behavior is proportional to
$\ell_0/L$, independent of the exact nonlinear response of the
linkers.

For a calculation of network mechanics the average tension
$\bar\tau$ in a filament is more relevant than the midpoint
tension~\cite{Morse98}. $\bar\tau$ is found by averaging the tension
profile given by Eq.~(\ref{LMM_tensionprofile}) along the backbone
of the filament. The ratio $\bar\tau/\tau_0$ is shown in the inset
of Fig.~\ref{LinMedModel}. We find that over a broad range of
strains $\bar\tau= 3/2 \tau_0$. During the cross-over regime the
ratio exhibits a peak with an amplitude that depends on the exact
ratio of $K_{EM}$ and $k_{cl}$.

\section{Self-Consistent medium model}\label{Self-Consistent Approach}
The linear treatment of the effective medium breaks down at large
strains. The network, consisting of a collection of many HR's, will
exhibit nonlinear response when the cross-linkers start to get
extended into their nonlinear regime. Thus, it is no longer
realistic to assume that the effective medium, which should reflect
the network elasticity, remains linear. In this section we extend
our model by requiring that the elasticity of the background medium
\emph{self-consistently} represents the nonlinear elasticity of the
constituent HR's. The elasticity of the medium should therefore
depend on the density of filaments and on the elasticity of an HR
averaged over all orientations. Thus, we require the stiffness per
cross-link of the effective medium $K_{EM}$ to be determined by the
stiffness of an HR
\begin{equation}\label{SCM_SelfConReq}
    K_{EM}=\frac{\alpha}{n L} \frac{d \tau_0}{d\epsilon}.
\end{equation}
The proportionality constant $\alpha$ depends on the detailed
structure of the network. In
section~\ref{continuumelastictyargument} we derive an expression for
$\alpha$ in the continuum elastic limit. The midpoint tension in a
rod can be written down analogous to Eq.~(\ref{LMM_tensionWLC})
\begin{equation}
\label{SCM_TensionSC}
   \tau_0(\epsilon) = \frac{n}{L} \int^{\frac{L}{2}} \, dx'
   \,x'
   \int^\epsilon \, d\epsilon' \frac{k_{cl}(x' \epsilon') \frac{\alpha}{n L} \frac{d \tau}{d\epsilon}(\frac{x'
    \epsilon'}{L/2})}{k_{cl}(x' \epsilon')+\frac{\alpha}{n L} \frac{d \tau}{d\epsilon}(\frac{x'
    \epsilon'}{L/2})
   },
\end{equation}
where $k_{cl}(u)$ is the derivative of the force-extension relation
of the cross-linker. Note that we have applied the same
approximation as we did in Eq.~(\ref{LMM_tensionWLC}). However, we
expect this approximation to hold even better here, since the medium
stiffens strongly as well as the cross-links.
Eq.~(\ref{SCM_TensionSC}) can be simplified to the following
differential equation for $\tau_0(\epsilon)$
\begin{eqnarray}
\label{SCM_SelfConDE}
   2 \frac{d\tau_0}{d\epsilon}&+&\epsilon
\frac{d^2\tau_0}{d\epsilon^2} = \\
&&\left\{
\begin{array}{ll}
    \frac{n L}{4} \frac{k_{cl}(\epsilon L/2) \frac{\alpha}{Ln}\frac{d\tau_0}{d\epsilon}}{k_{cl}(\epsilon L/2)+\frac{\alpha}{Ln}
\frac{d\tau_0}{d\epsilon}} & \textrm{if $\epsilon < \frac{\ell_0}{L/2}$}\\
    \\
    \frac{\alpha}{4} \frac{d\tau_0}{d\epsilon}
     & \textrm{if $\epsilon \geq\frac{\ell_0}{L/2}$}
    \end{array} \right.\nonumber
\end{eqnarray}
We find the following behavior of the model with WLC cross-linkers:
Below the characteristic strain for nonlinear response
$\epsilon_c=2\ell_0/L$, the tension in a rod depends approximately
linearly on strain. This linearity will be reflected in the
self-consistent effective medium, and consequently, the model shows
behavior similar to the linear medium model up to the critical
strain. By solving Eq.~(\ref{SCM_SelfConDE}) we find the midpoint
tension $\tau_0$ in a rod as a function of extensional strain
$\epsilon$. Beyond the critical strain the tension depends highly
nonlinearly on strain, with a derivative that increases as
\begin{equation}\label{SCC_assymptotic}
     \frac{d\tau_0}{d\epsilon} \sim \epsilon^{\frac{\alpha}{4}-1}.
\end{equation}
Note that unlike in the linear medium model, where the derivative
asymptotes to a final value set by $K_{EM}$, here $d\tau_0/d\gamma$
increases indefinitely. For the HFE cross-linkers we find similar
behavior, although in that case the cross-over between the linear
regime and the asymptotic stiffening regime is more abrupt.
\subsection{Continuum elastic limit}\label{continuumelastictyargument}
Here we derive an expression for $\alpha$ in the continuum elastic
limit. Note that this will only be a good approximation for a dense,
isotropic network. The modulus of the medium $G_{EM}$ can be
expressed in terms of the stiffness $\frac{d \tau_0}{d \gamma}$ of a
HR by averaging over rod orientations~\cite{Gittes98}
\begin{equation}\label{SCM_Gnetwork}
    G_{network}=\frac{1}{15}\rho \frac{d \bar \tau}{d \gamma},
\end{equation}
where $\rho$ is the length of filament per unit volume. $\rho$ can
also be expressed in terms of the mesh-size $\rho=1/\xi^2$. In the
linear medium treatment in section~\ref{Linear medium model} we
found that $\bar\tau=\frac{2}{3}\tau_0$. Thus, the network modulus
reads
\begin{equation}\label{SCM_Gnetwork2}
    G_{network}=\frac{2}{45}\frac{1}{\xi_2} \frac{d\tau_0}{d
    \gamma}.
\end{equation}
We proceed by relating $K_{EM}$ to $G_{network}$, which enables us
to find an expression for $\alpha$. Consider a rigid rod of diameter
$a$ and length $L$, which we use as a microrheological probe in an
effective elastic medium with a shear modulus $G_{EM}$. If the rod
is displaced along its axis, it will induce a medium deformation
$\delta \ell$ that leads to a restoring force acting along its
backbone. The restoring force per unit length is given by $2 \pi
G_{EM}/\log(L/a)\times\delta\ell$. Here we ignore the log term,
which is of order $2 \pi$. Thus, the stiffness of the medium per
cross-link $K_{EM}$ is related to $G_{EM}$ by
\begin{equation}\label{SCM_GEM}
    K_{EM}=\frac{L}{n}G_{EM}.
\end{equation}
By requiring $G_{EM} = G_{network}$ we find $\alpha$ from
Eqns.~(\ref{SCM_Gnetwork2}) and (\ref{SCM_GEM})
\begin{equation}\label{SCM_alpha}
    \alpha=\frac{2}{45}\left(\frac{L}{\xi}\right)^2.
\end{equation}
Note that for a dense network $\alpha \gg 1$.
\section{3D Network calculation}\label{3D network calculation}
In this section we describe in detail how the macroscopic mechanical
properties of a uniformly deforming network can be inferred from
single filament properties. This procedure has been used to describe
the viscoelastic~\cite{Gittes98} and nonlinear elastic
properties~\cite{Gardel04,Storm05,Janmey07} of semiflexible polymer
networks with point-like rigid cross-links, although a detailed
derivation of this theory is still lacking. The main assumption of
this calculation is a uniform, or affine deformation of the network.
The validity of the affine treatment of cross-linked semiflexible
polymer networks has been subject to much debate. Interestingly, 2D
simulations in the zero temperature limit have found that the
deformation can be both affine and non-affine depending on the
density of the network and filament
rigidity~\cite{Head03,Wilhelm2003}. Here we derive the affine theory
for the case of a filamentous network with point-like rigid
cross-links. Then we show how this framework can be used together
with the effective medium approach to describe the mechanics of
stiff polymer networks with flexible cross-links.

Consider a segment of a filament between two cross-links with an
initial orientation $\hat{n}$. When subjected to a deformation
described by the Cauchy deformation tensor $\Lambda_{ij}$, this
filament segment experiences an extensional strain directed along
its backbone
\begin{equation}\label{NL_extensionalstrain}
\epsilon=|\Lambda \hat{n}|-1.
\end{equation}
This extensional strain leads either to compression or extension in
the polymer segment depending on its orientation, and thus results
in a tension $\tau(|\Lambda \hat{n}|-1)$. The contribution of this
tension to the macroscopic stress depends also on the orientation of
the polymer segment. By integrating over contributions of the
tension over all orientations accordingly, we can compute the
macroscopic stress tensor $\sigma_{ij}$. We calculate the
contribution of the tension in a polymer segment with an initial
orientation $\hat{n}$ as follows. The deformation $\Lambda_{ij}$
transforms the orientation of the segment into $n_j' =
\Lambda_{jk}n_k/|\Lambda \hat{n}|$. Thus, the length density of
polymers with an orientation $\hat{n}$ that cross the $j$-plane is
given by $\frac{\rho}{\text{det}\Lambda}\Lambda_{jk} n_k$, where the
factor $\text{det}\Lambda$ accounts for the volume change associated
with the deformation. For the network calculations in this article
we consider only simple shear, which conserves volume
($\text{det}\Lambda=1$). The tension in the $i$-direction in a
filament with an initial orientation $\hat{n}$, as it reorients
under strain, is $\tau(|\Lambda \hat{n}|-1)\Lambda_{il}n_l/|\Lambda
\hat{n}|$. Thus, the (symmetric) stress tensor reads~\cite{Storm05}
\begin{equation}\label{NL_stresstensor}
\sigma_{ij}=\frac{\rho}{\text{det}\Lambda}\left\langle\tau(|\Lambda
\hat{n}|-1)\frac{\Lambda_{il}n_l \Lambda_{jk}n_k}{|\Lambda
\hat{n}|}\right\rangle.
\end{equation}
The angular brackets indicate an average over the initial
orientation of the polymer chains.

One remarkable feature follows directly from
Eq.~(\ref{NL_stresstensor}). A nonlinear force extension curve for
the filaments is not strictly required for a nonlinear network
response~\cite{Conti}. To demonstrate this we express the
extensional strain of a filament explicitly in terms of the strain
tensor $\gamma_{kl}$
\begin{equation}\label{NL_extensionalstrain}
\epsilon=\sqrt{1+2 u_{kl}\hat{n}_k\hat{n}_l}-1.
\end{equation}
Thus the extensional strain of a filament depends nonlinearly on the
macroscopic strain of the network. Additionally, the reorientation
of the filaments under strain leads to an increasingly more
anisotropic filament distribution. Remarkably, these geometric
effects result in a stiffening of the shear modulus under shear
strains of order 1, even in the case of Hookean filaments. At large
strains all filaments are effectively oriented in the strain
direction, which limits the amount of stiffening to a factor of 4
(2D networks) and 5 (3D networks) over the linear modulus at strains
of order 10. Thus the stiffening due to this effect occurs only at
large strains and is limited to a factor 5. Therefore we expect this
mechanism to have a marginal contribution to the more dramatic
stiffening that is observed in biopolymer gels at strains
$<1$~\cite{Gardel04,Storm05}. We would like to stress that the
geometric stiffening discussed above has a different nature than the
geometric stiffening discussed by~\cite{Heussinger072,Onck2005}. In
their case, the stiffening is attributed to a cross-over between an
elastic response dominated by soft bending modes in the zero strain
limit and a stiffer stretching mode dominated regime at finite
strains. In the affine calculation described here, only stretching
modes are considered.

By limiting ourself to a small strain limit, we can exclude the
geometric stiffening effects discussed above. This is instructive,
since it allows us to study network stiffening due to filament
properties alone, and it is a very good approximation for most
networks since the nonlinear response typically sets in at strains
$<1$. For a volume conserving deformation ($\text{det}\Lambda = 1$)
in the the small strain limit the stress tensor in
Eq.~(\ref{NL_stresstensor}) reduces to~\cite{Gittes98}
\begin{equation}\label{NL_stresstensorSSlimit}
\sigma_{ij}=\rho
\left<\tau(\gamma_{kl}\hat{n}_k\hat{n}_l)\hat{n}_i\hat{n}_j\right>,
\end{equation}
In this limit the geometric stiffening mechanism discussed above is
absent. Next we show explicitly how to calculate the shear stress
$\sigma_{xz}$, in the $z$-plane for a network, which is sheared in
the $x$-direction. A filament segment with an orientation given by
the usual spherical coordinates $\theta$ and $\varphi$ undergoes an
extensional strain
\begin{eqnarray}\label{NL_strainangles}
    \epsilon &=& \sqrt{1+2\gamma \cos(\varphi) \sin(\theta)
   \cos(\theta)+\gamma^2 \cos^2(\theta)}-1 \nonumber\\
      &\approx&\gamma \cos(\varphi) \sin(\theta)
   \cos(\theta),
\end{eqnarray}
where we have used a small strain approximation in the second line.
The tension in this segment contributes to the $xz$-component of the
stress tensor through a geometric multiplication factor $\cos(\phi)
\sin(\theta) \cos(\theta)$, where the first two terms are due to a
projection of the forces in the $x$-direction and the second term is
due to a projection of the orientation of the filament into the
orientation of the $z$-plane. The stress in the $xz$-direction is
thus given by
\begin{eqnarray}\label{NL_stresstensorangles}
\sigma_{xz} &=& \frac{\rho}{4 \pi} \int_0^{\pi} \,\int_0^{2\pi} \,
d\theta d\varphi
   \, \sin(\theta)\big{\{}\\
 &\tau&\!\!\!\!\left[\gamma \cos(\varphi) \sin(\theta) \cos(\theta)\right] \cos(\varphi) \sin(\theta)
   \cos(\theta)\big{\}}.\nonumber
\end{eqnarray}
Since we limit ourselves to the small strain limit, we do not
account for a redistribution of the filament orientations by the
shear transformation in this equation.
\subsection{semifexible polymer networks with rigid point-like cross-links}
In this section we show how the affine framework can be used to
compute the elastic response of a network with inextensible
semiflexible polymers connected by point-like rigid cross-links.

Consider a segment of an inextensible semiflexible polymer of length
$\ell_c$ between two rigid cross-links in the network. Thermal
energy induces undulations in the filament, which can be stretched
out by an applied tension. By adopting the WLC model in the
semiflexible limit $\ell_c \gtrsim \ell_p$, the force-extension
relation of this segment has been shown to be given implicitly
by~\cite{MacKPRL95}
\begin{equation}\label{NLrigidlinker_forceextenion}
\delta \ell=\frac{\ell_c^2}{\pi^2 \ell_p}\sum_{n=1}^\infty
\frac{\phi}{n^2(n^2+\phi)},
\end{equation}
where $\phi$ is the tension $\tau$ normalized by the buckling force
threshold $\kappa \frac{\pi^2}{\ell_c^2}$. This relationship can be
inverted to obtain the tension as a function of the extension $\delta\ell$:
\begin{equation}
\tau=\kappa \frac{\pi^2}{\ell_c^2}\phi\left(\delta\ell/\delta\ell_{\mbox{\rm\scriptsize max}}\right),
\end{equation}
where $\delta\ell_{\mbox{\rm\scriptsize max}}=\frac{1}{6}\ell_c^2/\ell_p$ is
the total stored length due to equilibrium fluctuations. This is also the maximum
extension, which can be found from
Eq.~(\ref{NLrigidlinker_forceextenion}) as
$\phi\rightarrow\infty$.
For small extensions
$\delta \ell$ this reduces to
\begin{equation}\label{NLrigidlinker_linearbehavior}
\tau=90 \frac{\kappa^2}{k_B T \ell_c^4} \delta\ell.
\end{equation}

This result can be inserted into Eq.~(\ref{NL_stresstensorSSlimit})
to find the linear modulus of the network
\begin{equation}\label{NLrigidlinker_linearmodulus}
G_0=6 \rho \frac{\kappa^2}{k_B T \ell_c^3}.
\end{equation}
For a network in either two or three dimensions, the maximally
strained filaments under shear are oriented at a 45 degree angle
with respect to the shear plane, meaning that the maximum shear
strain is
\begin{equation}\label{NLrigidlinker_maxstrain}
\gamma_{\mbox{\rm\scriptsize max}}=\frac{1}{3}\frac{\ell_c}{\ell_p}.
\end{equation}
Using the small strain approximation(as in
Eq.~(\ref{NL_stresstensorSSlimit})), we can calculate the nonlinear
network response
\begin{eqnarray}\label{NLrigidlinker_stresstensorangles}
   \frac{\sigma}{\sigma_c} &=& \frac{1}{4 \pi} \int_0^{\pi} \,\int_0^{2\pi} \, d\theta d\varphi
   \, \sin(\theta)\big{\{}\\
   &\phi&\!\!\!\!\left[\tilde{\gamma} \cos(\varphi) \sin(\theta) \cos(\theta)\right] \cos(\varphi) \sin(\theta)
   \cos(\theta)\big{\}}\nonumber
\end{eqnarray}
where we define the critical stress to be $\sigma_c=\rho \frac{\kappa}{\ell_c^2}$. We have also
defined $\tilde{\gamma}=\gamma/\gamma_c$, where the critical strain for the network is
\begin{equation}\label{NLrigidlinker_critstrain}
\gamma_c=\frac{1}{6}\frac{\ell_c}{\ell_p}.
\end{equation}
This equation demonstrates that
the nonlinear response of a network of inextensible semiflexible
polymers with rigid cross-links is universal~\cite{Gardel04} for small strains. Note,
however, that this would not hold if we would use the full nonlinear
theory from Eq.~(\ref{NL_stresstensor}), valid for arbitrarily large
strains. Thus, geometric stiffening effects may lead to small
departures from universality. Alternatively, universality may break
down as a result of enthalpic stretching of the polymer
backbone~\cite{Storm05}.

In this section we have assumed that at zero strain all filament
segments are at their equilibrium zero-force length. However,
cross-linking of thermally fluctuating polymers will result in
cross-linking distances both smaller and greater than their
equilibrium length. This effect, which is ignored in our discussion
here, leads to internal stresses build into the network during the
gelation~\cite{Storm05}.

The universal nonlinear elastic response for a semiflexible polymer
network with rigid cross-links is shown in Figs.~\ref{SC_Kvsstrain}
and \ref{SC_Kvsstress}. The divergence of the differential modulus
beyond the critical strain is of the form
$\sim\frac{1}{(1-\gamma_{\mbox{\rm\scriptsize max}})^2}$, as
depicted in Fig.~\ref{SC_Kvsstrain}. This results into a powerlaw
stiffening regime of the form $K \sim \sigma^{3/2}$, as shown in the
inset of Fig.~\ref{SC_Kvsstress}. This prediction is consistent with
experiments on actin gels with the rigid cross-linker
scruin~\cite{Gardel04}.

\subsection{stiff polymer networks with highly flexible
cross-links} For a network with flexible cross-links we do not
consider the tension in filament segments, but rather the average
tension $\bar{\tau}$ in the whole filament. By using the effective
medium approach we can compute the average tension in a filament as
a function of the orientation of the rod and the macroscopic shear
strain $\gamma$. Contributions to the stress from the average
tension in the rods are integrated over all orientations according
to Eq.~(\ref{NL_stresstensorangles}). In our description  we thus
assume affine deformation of the network on length scales $>L$.
Note, however, that we do not assume that the cross-links deform
affinely.

\begin{figure}
\centering
\includegraphics[width=240 pt]{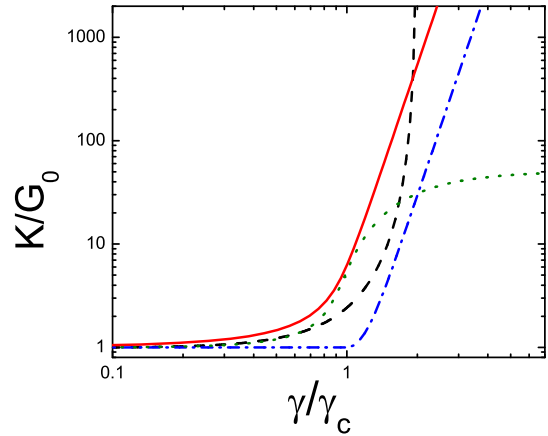}
\caption{(Color online) The differential modulus $K
=d\sigma//d\gamma $ normalized by the linear modulus $G_0$ as a
function of strain normalized by the critical strain $\gamma_c$. The
universal curve for a semiflexble polymer network with rigid
cross-links is shown as a black dashed curve. We also show the
results of the self-consistent model with WLC cross-links (red solid
curve) and simple cross-links (blue dash-dotted curve), the linear
medium model with WLC cross-links with $K_{EM} = 100 k_{cl}$ (green
dotted curve)} \label{SC_Kvsstrain}
\end{figure}

\begin{figure}
\centering
\includegraphics[width=240 pt]{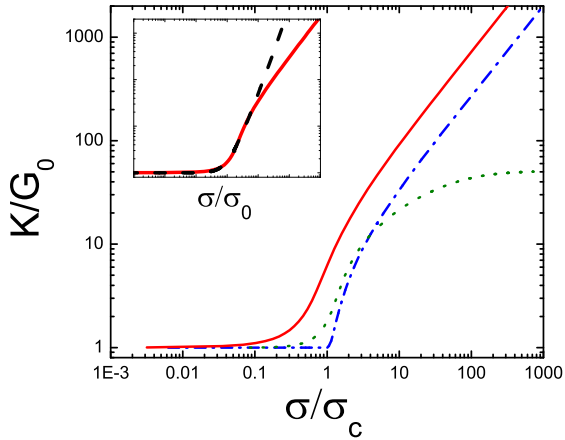}
\caption{(Color online) The differential modulus
$K=d\sigma//d\gamma$ normalized by the linear modulus Go as a
function of stress normalized by the critical stress $\sigma_c$ for
the self-consistent model with WLC cross-links (red solid curve),
HFE cross-links (blue dash-dotted curve) and the linear medium model
with WLC cross-links with $K_{EM} = 100 k_{cl}$ (green dotted
curve). The inset shows the rigid linker model together with the
self-consistent model for a network with flexible cross-links. In
this case the stress is normalized by a stress $\sigma_0$, which
marks the knee of the curve.} \label{SC_Kvsstress}
\end{figure}

We find both from the linear medium model and the self-consistent
model for a network with highly flexible cross-links that the linear
modulus is approximately given by
\begin{equation}\label{3D_linmod}
     G_0\approx\frac{1}{8}\rho n k_{cl} L.
\end{equation}
The appearance of the filament length $L$ in this equation is
remarkable, and is due to the non-uniform deformation profile of the
cross-links, which enhances the forces applied by the cross-links
further from the midpoint of the filament. The onset of nonlinear
elastic response occurs at a critical strain
\begin{equation}\label{3D_critstrain}
     \gamma_c=4 \frac{\ell_0}{L}.
\end{equation}
The full nonlinear response as predicted by our model is shown in
Figs.~\ref{SC_Kvsstrain} and ~\ref{SC_Kvsstress}. The results of the
linear medium model with WLC cross-links, as shown with a green
dotted line, are qualitatively similar to the results of the 1D
model(see Fig.~\ref{LinMedModel}). For the self-consistent model we
find that beyond $\gamma_c$ the differential modulus increases as a
powerlaw, as shown in Fig.~\ref{SC_Kvsstrain}. Interestingly, we
find only a small quantitative difference between the model with HFE
and WLC cross-links.

The differential modulus $K=d\sigma/d\gamma$ is plotted as a
function of stress in Fig.~\ref{SC_Kvsstress}. The stress is
normalized by the critical stress $\sigma_c$, which we define here
as
\begin{equation}\label{3D_critstress}
     \sigma_c=G_0 \gamma_c=\frac{1}{2}\rho n k_{cl} \ell_0.
\end{equation}
We find a sharp increase in stiffness beyond the
critical stress, which quickly asymptotes to a
powerlaw regime, where the exponent is given by
$1-1/(\frac{1}{60}(L/\xi)^2-1)$. Interestingly,
this exponent does not depend on the exact form
of the nonlinear response of the cross-linkers.
This exponent emerges as a consequence of the
finite extendability of the cross-links and the
non-uniform deformation profile along the
backbone of the filament. Remarkably, the
powerlaw exponent is not universal. However, in
the dense limit we consider in our model, the
deviation to an exponent of 1 is $\ll 1$ and
depends only weakly on the ratio $L/\xi$. As an
example, we consider a typical \emph{in vitro}
network for which $\xi = 0.3 \mu \text{m}$ and
the average filament length is $L = 15 \mu
\text{m}$. For this case we find an exponent of
$0.98$. The asymptotic powerlaw regime with an
exponent $\approx 1$, as predicted by our model
is consistent with recent experimental data on
actin networks cross-linked by
filamin~\cite{Gardel06,Karen}.

The inset of Fig.~\ref{SC_Kvsstress} shows the rigid linker model
together with the self-consistent model for a network with flexible
cross-links. In this case the stress is normalized by a stress
$\sigma_0$, which marks the knee of the curve.

\section{Tension profiles and single cross-linker force estimate}
Recently, there has been much debate on the mechanical response of
actin binding proteins such as filamin. Specifically, it is
discussed whether the cross-links stiffen, unfold or unbind under
tension in both physiological or \emph{in vitro} conditions. This
issue has major implications for the dynamical and mechanical
properties of the cytoskeleton. The discussion has been partially
resolved recently by single molecule~\cite{Ferrer08} and bulk
rheology~\cite{Karen} experiments on the actin-filamin system. These
experiments indicate that cross-links unbind at forces well below
the force required for domain unfolding. It is crucial for the bulk
rheology experiment, to be able to infer the forces experienced by a
single cross-linker from the measured mechanical stress. In this
section we show that by using the shape of the tension profile, we
can relate a macroscopic quantity such as the stress to the maximum
force experienced by a single cross-linker in the network.

The tension along a single filament is not uniform in networks of
stiff finite length filaments and incompliant
cross-links~\cite{Head03,Heussinger07}. It was found in simulations
that in the affine regime the tension profile is flat close to the
midpoint and the tension decreases exponentially towards the
boundaries of the filament. In the non-affine regime a different
tension profile has been reported, in which the tension decreases
linearly towards the ends~\cite{Heussinger07}. In the case of a
flexibly cross-linked network of stiff polymers we also expect a
non-uniform tension profile, although in this case the underlying
physics is different. The deformation of a cross-linker at a
distance $x$ from the midpoint of the rod is $u_{cl}\sim x\gamma$
and, consequentially, cross-links further away from the midpoint
exert larger forces on the rod, resulting in a non-uniform tension
profile.

\begin{figure}
\centering
\includegraphics[width=240 pt]{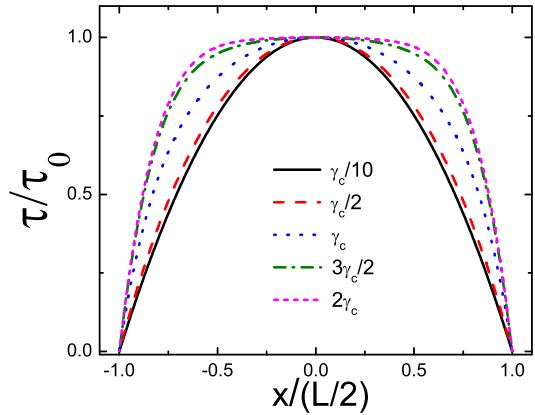}
\caption{(Color online) The reduced tension profile along the rod,
normalized by the midpoint tension $\tau_0$. This profile is
calculated with the self-consistent model with WLC cross-links.}
\label{tensionprofile}
\end{figure}

We can calculate the tension profile for a given rod using
Eq.~(\ref{LMM_tensionprofile}). In the limit of highly flexible
cross-linkers, the tension profile in the linear elastic regime is
given by
\begin{equation}\label{TP_linmedmodelprofile}
     \tau(\epsilon,x)= \frac{n}{L} \frac{k_{cl} K_{EM}}{k_{cl}+K_{EM}}\frac{1}{2}
     \left(x^2-\left(\frac{L}{2}\right)^2\right)\epsilon.
\end{equation}
The tension profiles as computed with the self-consistent model with
WLC cross-links are shown for various strains in
Fig.~\ref{tensionprofile}. For low strains we find a parabolic
profile, which flattens out towards the edges for larger strains.

We now proceed to estimate the force experienced by a single
cross-linker. For an affinely deforming network in the linear
response regime Eq.~(\ref{NL_stresstensor}) simplifies to
\begin{equation}\label{TP_stress}
     \sigma=\frac{1}{15}\rho\bar{\tau}(\gamma).
\end{equation}
Filaments at a $45^\circ$ angle with respect to the stress plane
bear the largest tension $\bar{\tau}_{max}$ and experience a strain
along their backbone of $\gamma/2$. Assuming linear response we find
$\bar{\tau}_{max}(\gamma)=\bar{\tau}(\gamma)/2$. In the case of a
parabolic tension profile, the average tension $\bar{\tau}$ in a
filament is related to the largest force $f_o$ experienced by a
cross-linker at the boundary of the rod by $\bar{\tau}=\frac{1}{6} n
f_0$. Thus, we can express the macroscopic stress in terms of the
maximum forces experienced by cross-linkers on the filaments under
the greatest load
\begin{equation}\label{TP_forceestimate}
     \sigma=\frac{1}{45}\rho n f_{max}.
\end{equation}
For the derivation of this equation we have assumed to be in the
linear response regime. In the nonlinear regime we expect the
expression to still hold approximately, although the prefactors will
change.

Kasza et al.~\cite{Karen} found that the failure stress of the
network $\sigma_{max}$ is proportional to the number of cross-links
per filament $n$ in actin networks with the flexible cross-linker
filamin. This suggests that filamin failure, rather than rupture of
single actin filaments is the cause for network breakage. In
contrast, for actin networks with the rigid cross-linker scruin,
which binds more strongly to actin than filamin, rupture of actin
was found to be the mechanism for network failure~\cite{Gardel04}.
On the basis of our model and the experimental data from
Ref.~\cite{Karen} we estimate filamin failure forces of order $1 -
5~\text{pN}$, far below the unfolding force $100~\text{pN}$. This
suggests that network failure is due to filamin unbinding. This is
consistent with recent single molecule experiments, which show that
filamin unbinding is favored over unfolding of the Ig-domains for
low loading rates~\cite{Ferrer08}.

\section{Implications and Discussion}\label{Discussion}
We have studied the nonlinear elasticity of stiff polymer network
with highly flexible cross-links. We find that the mechanics of such
a network is controlled by network connectivity expressed in the
number of cross-links per filament $n$. This was found earlier in
experiments on actin-filamin gels~\cite{Karen}, providing strong
experimental evidence for cross-link dominated mechanics in these
networks. Within this picture, stiffening occurs at a strain where
the cross-links are stretched towards their full extension. As a
result, we expect $\gamma_c$ to be proportional to the molecular
weight of the cross-linker $\ell_0$. This prediction is consistent
with the results of Wagner et al.~\cite{Wagner06}, where cross-link
length was varied, while keeping the average filament length fixed.
Interestingly, they observed larger values of $\gamma_c$ than
expected either from our model or based on
Refs~\cite{Gardel06,Wagner06,Karen}.

In addition, we find here that the filament length $L$ plays an
important role in the nonlinear elasticity of these networks. In
particular, the onset of nonlinear response $\gamma_c\sim\ell_0/L$
depends crucially on filament length. This has been confirmed by
recent experiments on actin-filamin gels, showing an approximate
inverse dependence of the $\gamma_c$ on actin filament
length~\cite{Liu}. The sensitivity of network response to filament
length, both in experiments and in our model, appears to be one of
the hallmarks of actin-filamin networks. On the one hand, this may
explain the apparent difference between the critical strains
reported in Refs.~\cite{Gardel06,Wagner06,Karen}. On the other hand,
this also suggests that it may be more even more important in such
flexibly cross-linked networks to directly control and measure
the filament length distribution than for other
\emph{in vitro} actin studies~\cite{PaulGelsolin}. Our model does
not account for filament length polydispersity. A distribution in
filament length is expected to smooth somewhat the sharp stiffening
transition predicted by our model.

The dependence of the critical strain for networks with flexible
cross-links observed in experiments and predicted by our model is in
striking contrast with the behavior found for rigidly cross-linked
networks. In the latter case theory predicts
$\gamma_c\sim\ell_p/\ell_c$ (see
Eq.~(\ref{NLrigidlinker_critstrain})), which is consistent with
experimental observations~\cite{Gardel04}. The insensitivity of the
nonlinear elasticity of dense networks cross-linked with rigid
linkers to filament length would suggest that network mechanics
cannot be effectively controlled by actin polymerization regulation.
We have shown here that the filament length plays a crucial role for
networks with flexible cross-links, which are abundant in the
cellular cytoskeleton. Thus regulating actin length by
binding/capping proteins such as gelsolin may enable the cell not
only to sensitively tune the linear elastic modulus, but also the
onset of the nonlinear response of its cytoskeleton.

In the nonlinear regime we expect the differential modulus to
increase linearly with stress for a dense flexibly cross-linked
network. This behavior is a direct consequence of the non-uniform
deformation profile along a filament and the finite extendability of
the cross-links, although it is independent of the exact shape of
the force-extension behavior of the cross-links. The powerlaw
stiffening $K \sim \sigma^y$ with $y \approx 1$ is consistent with
recent experiments on actin-filamin gels~\cite{Gardel06,Karen}. This
stiffening behavior is very different from the nonlinear response
observed for actin gels with rigid cross-links for which a powerlaw
exponent of $3/2$ is observed~\cite{Gardel04}, consistent with
theory for an affine response governed by the stretching out of
thermal fluctuations of the actin filaments. Interestingly, \emph{in
vivo} experiments show that cells also exhibit power-law stiffening
with an exponent of 1~\cite{Fernandez}.

\begin{figure}
\centering
\includegraphics[width=240 pt]{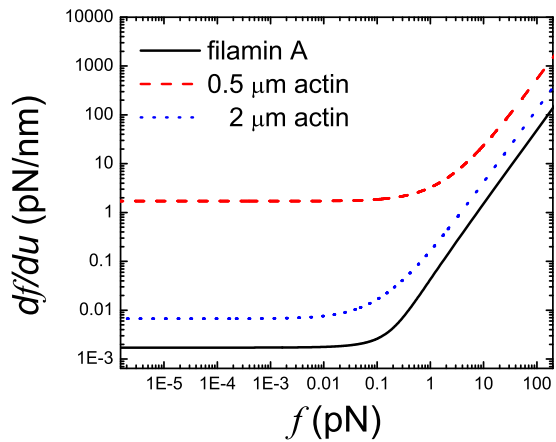}
\caption{(Color online) The differential stiffness $d \emph{f}/du$
as a function of force $\emph{f}$ for a filamin cross-linker (solid
line) and for several F-actin polymer segment lengths.}
\label{stiffnesscomparison}
\end{figure}

In this article we have examined a limit in which the stiffness of
the cross-links is small compared to the stiffness of an F-actin
segment between adjacent cross-links. For a large flexible
cross-linker such as filamin this is clearly a good approximation in
the linear regime. However, as the cross-links stiffen strongly they
could, in principle, become as stiff as the actin segment. This
would have a dramatic consequence for the nonlinear response of the
network. To investigate this we have calculated the differential
stiffness $d\emph{f}/du$ as a function of force \emph{f} for a
filamin cross-linker and an actin segment with a length $0.5~
\text{to}~ 2~\ \mu\text{m}$, spanning the range of typical distances
between cross-links in dense and sparse networks respectively. This
result is shown in Fig.~\ref{stiffnesscomparison}. We find the
differential stiffness of a filamin cross-link is always smaller
than for an F-actin segment, even at large forces in the nonlinear
regime. This justifies our approach, in which we have ignored the
compliance of the actin, for a broad range of experimentally
accessible polymer/cross-linking densities. However, at sufficiently
high filamin concentrations, it may be possible that individual
network nodes involve multiple cross-linkers, in which case the
actin filament compliance may also become relevant. Thus the affect
of the compliance of F-actin remains an interesting topic for
further research.

We also use our model to study these networks on a more microscopic
level, such as the non-uniform tension profiles along the filament
backbone. These profiles can be used to establish a relation between
the macroscopic stress and the largest force experienced by a single
cross-linker in the network. This allows us to estimate the forces
experienced by filamin cross-links under typical \emph{in vitro} and
\emph{in vivo} conditions. We find that the load on these
cross-links is not sufficiently high to lead to significant domain
unfolding of the filamin Ig-domains, even at stresses large enough
to rupture the network. Indeed both rheology experiments on actin
filamin gels and single molecule experiments indicate that unbinding
occurs well before domain unfolding.

In other large flexible cross-links such as
spectrin~\cite{Rief1999}, domain unfolding occurs at lower, more
relevant forces. In this case the domain unfolding could have a
dramatic affect on the nonlinear viscoelasticity of such networks.
In previous work, DiDonna and Levine have simulated 2D cross-linked
networks, where they have assumed a sawtooth force-extension curve
for the cross-linkers to mimic domain unfolding~\cite{DiDonna06}.
They report a fragile state with shear softening when an appreciable
number of cross-linkers are at the threshold of domain unfolding.
Our model is based on the stiffening of the cross-linkers, which
occurs at forces far below those required for domain unfolding.
This leads to strain stiffening at a point where only a fraction of
cross-linkers are at their threshold for nonlinear response. Thus in
both our model and that of Ref.~\cite{DiDonna06} the network
responds strongly to small strain changes, though in an opposite
manner: stiffening in the present case vs softening in
Ref.~\cite{DiDonna06}.

In related work, Dalhaimer, Discher, and Lubensky show that
isotropic networks linked by large compliant cross-linkers exhibit a
shear induced ordering transition to a nematic
phase~\cite{DischerLubensky}. It would be interesting to investigate
the affect of the nonlinear behavior of the cross-links on this
transition. In the present calculation we have assumed an isotropic
network. An ordering transition, which results in a strong alignment
of filaments will dramatically affect the nonlinear elasticity of
the network.

In this article we have studied networks of stiff polymers linked by
highly flexible cross-links. Both experiments~\cite{Gardel06,Karen}
and our model~\cite{broedersz08} indicate that these networks have
novel nonlinear rheological properties. We find that the network
mechanics is highly tunable. By varying filament length,
cross-linker length and network connectivity we can sensitively
regulate the linear and nonlinear elasticity over orders of
magnitude. These unique properties can be exploited in the design of
novel synthetic materials.
\begin{acknowledgments}
We thank K.\ Kasza, G.\ Koenderink, E.\ Conti and M.\ Das for
helpful discussions. This work was funded in part by FOM/NWO.
\end{acknowledgments}

\end{document}